\documentclass{article}

 \usepackage{amsmath}
 \usepackage{tikz}
 \usepackage{placeins}
 \usetikzlibrary{positioning, shapes.geometric, arrows.meta, fit, calc}

 \usepackage[dblblindworkshop, final]{neurips_2025}

\usepackage[utf8]{inputenc} 
\usepackage[T1]{fontenc}    
\usepackage{hyperref}       
\usepackage{url}            
\usepackage{booktabs}       
\usepackage{amsfonts}       
\usepackage{nicefrac}       
\usepackage{microtype}      
\usepackage{xcolor}         

\title{\texttt{FlowLensing}: Simulating Gravitational Lensing with Flow Matching}
\workshoptitle{Machine Learning and the Physical Sciences}

\author{%
Hamees Sayed$^{1,2}$ \quad Pranath Reddy$^{3}$ \quad Michael W. Toomey$^{4}$ \quad Sergei Gleyzer$^{5}$\\[4pt]
$^{1}$Smallest AI \quad
$^{2}$Indian Institute of Technology Madras, India \\
$^{3}$University of Florida, Gainesville, FL 32611, USA \\
$^{4}$Center for Theoretical Physics, Massachusetts Institute of Technology, Cambridge, MA 02139, USA \\
$^{5}$Department of Physics and Astronomy, University of Alabama, Tuscaloosa, AL 35401, USA\\[6pt]
\texttt{hamees@smallest.ai} \quad
\texttt{kumbam.pranath@gmail.com} \quad
\texttt{mtoomey@mit.edu} \quad
\texttt{sgleyzer@ua.edu}
}

\begin{document}

\maketitle

\begin{abstract}
    Gravitational lensing is one of the most powerful probes of dark matter, yet creating high-fidelity lensed images at scale remains a bottleneck. Existing tools rely on ray-tracing or forward-modeling pipelines that, while precise, are prohibitively slow. We introduce \texttt{FlowLensing}, a Diffusion Transformer-based compact and efficient flow-matching model for strong gravitational lensing simulation. \texttt{FlowLensing} operates in both discrete and continuous regimes, handling classes such as different dark matter models as well as continuous model parameters ensuring physical consistency. By enabling scalable simulations, our model can advance dark matter studies, specifically for probing dark matter substructure in cosmological surveys. We find that our model achieves a speedup of over 200$\times$ compared to classical simulators for intensive dark matter models, with high fidelity and low inference latency. \texttt{FlowLensing} enables rapid, scalable, and physically consistent image synthesis, offering a practical alternative to traditional forward-modeling pipelines.
\end{abstract}

\section{Introduction}
Gravitational lensing \cite{Bartelmann_2010} occurs when light from a distant galaxy or quasar is bent by the gravity of an intervening object, typically by a galaxy and its more massive dark matter halo. This phenomenon distorts and magnifies the background source, sometimes producing multiple images, and has become a powerful tool to probe the distribution of dark matter and test competing particle models.

Simulating realistic lensing images at scale, however, is computationally demanding. Existing tools like \texttt{lenstronomy} \cite{Birrer_2018} and \texttt{PyAutoLens} \cite{10.1093/mnras/sty1264} can generate high fidelity images by solving the lens equation via ray tracing or grid-based solvers, but their high cost for complex dark matter models makes them unsuitable for large statistical studies -- in particular when attempting to study affects from dark matter substructure. Deep learning based generative models offer speedups, yet current approaches often struggle with fidelity, conditioning control, or slow inference due to long sampling chains.

To address these challenges, we propose \texttt{FlowLensing}, a flow-matching \cite{lipman2023flow} model with diffusion transformer (DiT) \cite{Peebles2022DiT} as the backbone that learns a direct mapping from astrophysical parameters to lensed images. Flow matching avoids iterative denoising, enabling faster and more stable sampling. Our method captures both broad dark matter scenarios and fine-grained lens properties, producing sharper, physically consistent images. As further discussed in Section~\ref{sec:downstream_eval}, strong classification and regression results serve as indirect evidence of physics consistency. Overall, our approach dramatically reduces inference time, making it a practical alternative to classical simulators.


\section{Datasets}

To train \texttt{FlowLensing}, we used two simulated datasets of gravitational lensing images, generated with \texttt{lenstronomy} to mimic Euclid-like survey observations. These datasets also capture dark matter substructure effects enabling tests of physical consistency and model performance. All images are 64$\times$64 pixels and normalized to $[-1, 1]$ for training. Details are provided in the subsections below.


\subsection{Dark Matter Model Conditioned (Discrete)}  
\label{subsec:discrete_class}

The first dataset is composed of simulated galaxy-galaxy strong lensing images that are generated using the publicly available simulation pipeline \texttt{lenstronomy}. There are 89,104 elements in the dataset, where every image is 64 x 64 pixels and is meant to resemble an observation typical of a Euclid-like survey. Furthermore, each simulated host lens is modeled as a sheared isothermal ellipse \cite{10.1093/mnras/stab930}, while each source is described via a Sersic light profile \cite{sersiclightprofile}. Every image within the dataset falls into one of three categories, each defined by a different dark matter substructure class. The first class is a baseline that involves no simulated substructure and has only a CDM \cite{Weinberg_2015} host halo as the deflector, while the second class assumes CDM substructure modeled as truncated NFW haloes \cite{10.1093/mnras/stad3894}. Finally, the third class models axionic dark matter \cite{OHare:2024nmr} with $m \approx 10^{-23}$ eV as vortex-like defects.

\subsection{Lensing Model Parameters (Continuous)}  
\label{subsec:continuous_class}

The second dataset has 30,000 elements and consists of only CDM images, with and without substructure, that were produced in the same way as the first dataset. So while all lensing simulations consist of the same dark matter type, a set of continuous parameters of the lens-source system were regressively extracted to define each one:
einstein radius ($\theta_E$), the coordinates of the source in the image plane with respect to the center of the image ($x,y$), and the slope of the subhalo mass function ($\beta$). Conditioning each simulated image in this way will allow for a more physically consistent performance from the flow matching model \cite{wildberger2023flow}.

\section{Methodology}

\subsection{Flow Matching with Diffusion Transformer}

Flow matching is a continuous-time generative modeling process that learns to transform samples from a simple prior distribution (typically Gaussian noise) to a target data distribution through a continuous flow. Unlike traditional diffusion models that rely on a fixed noising schedule, flow matching directly learns the vector field that guides the transformation process. Compared to score-based diffusion methods \cite{song2021scorebased}, flow matching avoids the need for stochastic differential equations and instead directly estimates the velocity field that transports particles along deterministic paths.

Given a pair of data $x_1 \sim p_{\text{data}}$ and noise $x_0 \sim \mathcal{N}(0, I)$, the interpolant at time $t \in [0, 1]$ is defined as
\[
x_t = (1-t) x_0 + t x_1.
\]
The corresponding optimal target velocity is
\[
v(x_t, t) = x_1 - x_0,
\]
and the training objective is to minimize the mean squared error between the predicted velocity $v_\theta(x_t, t, c)$ and the ground truth $v(x_t, t)$:
\[
\mathcal{L}(\theta) = \mathbb{E}_{x_0, x_1, t, c} \left[ \| v_\theta(x_t, t, c) - (x_1 - x_0) \|^2 \right],
\]
where $c$ represents the conditioning signal.

We implement $v_\theta$ using a DiT, which operates on image patches with self-attention \cite{10.5555/3295222.3295349} and integrates conditioning information through adaptive layer normalization (adaLN) \cite{DBLP:journals/jccee/GuoWYML22}. To improve controllability, we apply classifier-free guidance \cite{ho2021classifierfree} by randomly dropping conditioning during training with probability $p_{drop}=0.1$. At inference, conditional and unconditional predictions are combined as
\[
\tilde{v}_\theta(x_t, t, y) = v_\theta(x_t, t, \emptyset) + w \cdot (v_\theta(x_t, t, y) - v_\theta(x_t, t, \emptyset)),
\]
where $w$ is the guidance weight and $\emptyset$ denotes the unconditional case.

\begin{figure}[t]
  \centering
  \includegraphics[width=0.9\textwidth]{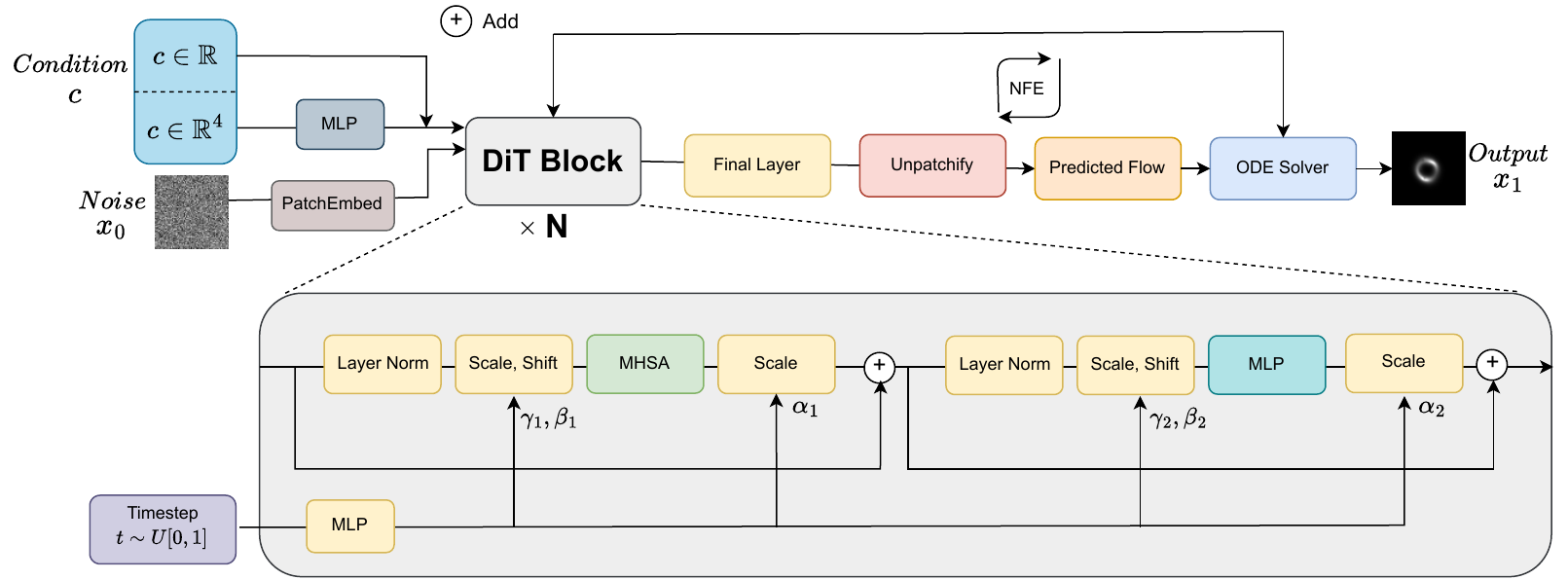}
  \caption{Schematic of \texttt{FlowLensing} inference.}
  \label{fig:architecture}
\end{figure}

\subsection{Experiment 1: Dark Matter Model Conditioned Generation}

In our first experiment, we condition \texttt{FlowLensing} on discrete classes representing different dark matter models. The conditioning variable $c$ takes one of three categorical values (CDM,axion,no substructure) mentioned in Section~\ref{subsec:discrete_class}. The conditioning is implemented through a learned embedding table that maps discrete class labels to the model's hidden dimension. During training, we apply classifier-free guidance by randomly replacing class labels with a special \emph{`null'} token, enabling unconditional generation and guidance during sampling.

\subsection{Experiment 2: Lensing Parameter Conditioned Generation}

This experiment extends conditioning to continuous lensing parameters for fine-grained control and interpolation. The conditioning vector $c \in \mathbb{R}^4$ is projected into the model’s hidden space using a multi-layer perceptron (MLP). Classifier-free guidance is applied by randomly masking the conditioning vector to zeros during training, with a learned null embedding used for unconditional cases. This setup enables the model to smoothly traverse parameter space and generate physically plausible lensing systems.

\section{Results}

We evaluate \texttt{FlowLensing} on two simulated datasets: one with discrete dark matter classes (Section~\ref{subsec:discrete_class}) and another with continuous lensing parameters (Section~\ref{subsec:continuous_class}). Performance is assessed using classification metrics (AUC) for discrete classes and regression metrics ($R^2$) for continuous parameters, alongside image quality metrics to quantify reconstruction fidelity. We also benchmark inference latency to highlight computational efficiency. All evaluations use a 30M-parameter model with a classifier-free guidance (CFG) scale of 2 and 100 denoising steps via an Euler ODE solver.


\subsection{Image Quality Metrics}

\begin{figure}[t]
    \centering   
    \begin{minipage}{0.8\textwidth}
        \centering
        \includegraphics[width=\linewidth]{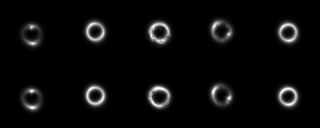}
        \caption{Real (top) vs. generated (bottom) images from \texttt{FlowLensing}.}
        \label{fig:real_vs_generated}
    \end{minipage}
\end{figure}

To quantitatively evaluate the fidelity of generated images, we report standard reconstruction and perceptual metrics. Mean Squared Error (MSE) measure pixel-level accuracy, Peak Signal-to-Noise Ratio (PSNR) captures overall signal quality, and Structural Similarity Index (SSIM) \cite{1284395} quantifies perceptual similarity. We additionally report Fréchet Inception Distance (FID) \cite{NIPS2017_8a1d6947} for completeness, although we note that the Inception model \cite{10.5555/3298023.3298188} used in FID is not trained on our domain-specific data and therefore may not be an ideal indicator of astrophysical realism. Inference efficiency is also critical: our model requires only 100 denoising steps versus 1000 for the baseline, achieving a $\sim$13.3$\times$ speedup. Table~\ref{tab:img-metrics} summarizes reconstruction quality and single-sample generation latency and in Figure~\ref{fig:real_vs_generated}, we provide a side-by-side comparison of real and generated images to qualitatively assess fidelity.



\begin{table}[ht]
\centering
\setlength{\tabcolsep}{10pt} 
\renewcommand{\arraystretch}{1.2} 
\begin{tabular}{lccccc}
\toprule
Model & MSE $\downarrow$ & FID $\downarrow$ & Latency (s) $\downarrow$ & PSNR $\uparrow$ & SSIM $\uparrow$ \\
\midrule
Ours & \textbf{0.0108} & \textbf{1.614} & \textbf{0.36} & \textbf{68.68} & \textbf{0.9993} \\
Baseline & 0.0110 & 87.312 & 4.8 & 30.78 & 0.8870 \\
\bottomrule
\end{tabular}
\renewcommand{\arraystretch}{1.0} 
\vspace{0.5em}
\caption{Comparison of reconstruction quality metrics against the baseline.}
\label{tab:img-metrics}
\end{table}



The baseline model for comparison is a DDPM \cite{10.5555/3495724.3496298} with a U-Net \cite{DBLP:journals/jccee/GuoWYML22} backbone, evaluated with 1000 NFE steps at inference, whereas our method achieves competitive performance using only 100 NFE steps. This highlights both the efficiency and effectiveness of our approach. 

\subsection{Downstream Evaluation: Classification and Regression}
\label{sec:downstream_eval}

To evaluate the utility of the learned representations, we assess two downstream tasks: classification and regression. For classification, a ResNet18 \cite{he2016residual} classifier trained on the original simulated dataset is evaluated on images produced by our model. For regression, the final fully connected layer of ResNet18 is replaced with a 4-dimensional linear layer to predict astrophysical parameters, serving primarily as a sanity check for parameter recoverability. 
The strong results across both tasks provide indirect evidence that our model preserves underlying physical consistency. The outcomes are summarized in Table~\ref{tab:downstream_results}.

\begin{table}[ht]
    \centering
    \setlength{\tabcolsep}{6pt} 
    \renewcommand{\arraystretch}{1.2} 
    \begin{tabular}{lcc|lccc}
        \toprule
        \multicolumn{3}{c|}{\textbf{Class. (AUC)}} & \multicolumn{4}{c}{\textbf{Reg. ($R^2$)}} \\
        \midrule
        \textbf{Class} & \textbf{Ours} & \textbf{Base.} & $\boldsymbol{x}$ & $\boldsymbol{y}$ & $\boldsymbol{\beta}$ & $\boldsymbol{\theta_E}$ \\
        \midrule
        CDM & \textbf{1.00} & 0.92 & 0.945 & 0.940 & 0.833 & Constant (1.281) \\
        Axion & \textbf{1.00} & 0.91 & & & & \\
        No Substructure & \textbf{1.00} & 0.75 & & & & \\
        \bottomrule
    \end{tabular}
    \renewcommand{\arraystretch}{1.0} 
    \vspace{0.5em}
    \caption{Downstream evaluation: classification (AUC) and regression ($R^2$).}
    \label{tab:downstream_results}
\end{table}

\section{Conclusion and Future Work}

Our work introduced \texttt{FlowLensing}, a flow-matching model with a diffusion transformer backbone that generated high-fidelity gravitational lensing images over 200$\times$ faster than traditional simulators (0.36s vs. 4.8s per sample). Conditioned on dark matter models (CDM, axion, no substructure) and parameters like the subhalo mass function slope $\beta$, it achieved superior image quality (PSNR: 68.68, SSIM: 0.9993) and physical accuracy, with perfect classification AUC (1.00) and strong regression $R^2$ scores (0.833--0.945; Section~\ref{sec:downstream_eval}). By enabling scalable, realistic simulations, \texttt{FlowLensing} advances dark matter studies, recovering $\beta$ to distinguish CDM from axion models and probe small-scale structures in surveys like Euclid.

Looking ahead, we aim to weave lensing equations into the model’s architecture for deeper physical fidelity, reducing reliance on post hoc checks. We also plan to benchmark \texttt{FlowLensing} against GANs and VAEs to explore its strengths across generative approaches.

\section{Acknowledgements}

We acknowledge useful conversations with Pranath Reddy. H.S. was a participant in the Google Summer of Code 2025 program. S.G. was supported in part by U.S. National Science Foundation award No. 2108645. Portions of this work were conducted in MIT’s Center for Theoretical Physics and partially supported by the U.S. Department of Energy under grant Contract Number DE-SC0012567. M.W.T is supported by the Simons Foundation (Grant Number 929255).



\bibliographystyle{plain}   

\begin{thebibliography}{10}

\bibitem{Bartelmann_2010}
Matthias Bartelmann.
\newblock Gravitational lensing.
\newblock {\em Classical and Quantum Gravity}, 27(23):233001, November 2010.

\bibitem{Birrer_2018}
Simon Birrer and Adam Amara.
\newblock lenstronomy: Multi-purpose gravitational lens modelling software package.
\newblock {\em Physics of the Dark Universe}, 22:189–201, December 2018.

\bibitem{sersiclightprofile}
Vincenzo Cardone.
\newblock The lensing properties of the sersic model.
\newblock {\em Astronomy and Astrophysics}, 415, 11 2003.

\bibitem{10.1093/mnras/stab930}
Matthew~R Gomer and Liliya L~R Williams.
\newblock Galaxy-lens determination of h0: the effect of the ellipse + shear modelling assumption.
\newblock {\em Monthly Notices of the Royal Astronomical Society}, 504(1):1340--1354, 04 2021.

\bibitem{DBLP:journals/jccee/GuoWYML22}
Yunhui Guo, Chaofeng Wang, Stella~X. Yu, Frank McKenna, and Kincho~H. Law.
\newblock Adaln: A vision transformer for multidomain learning and predisaster building information extraction from images.
\newblock {\em J. Comput. Civ. Eng.}, 36(5), 2022.

\bibitem{he2016residual}
Kaiming He, Xiangyu Zhang, Shaoqing Ren, and Jian Sun.
\newblock {Deep Residual Learning for Image Recognition}.
\newblock In {\em Proceedings of 2016 IEEE Conference on Computer Vision and Pattern Recognition}, CVPR '16, pages 770--778. IEEE, June 2016.

\bibitem{10.1093/mnras/stad3894}
Felix~M Heinze, Giulia Despali, and Ralf~S Klessen.
\newblock Not all subhaloes are created equal: modelling the diversity of subhalo density profiles in tng50.
\newblock {\em Monthly Notices of the Royal Astronomical Society}, 527(4):11996--12015, 12 2023.

\bibitem{NIPS2017_8a1d6947}
Martin Heusel, Hubert Ramsauer, Thomas Unterthiner, Bernhard Nessler, and Sepp Hochreiter.
\newblock Gans trained by a two time-scale update rule converge to a local nash equilibrium.
\newblock In I.~Guyon, U.~Von Luxburg, S.~Bengio, H.~Wallach, R.~Fergus, S.~Vishwanathan, and R.~Garnett, editors, {\em Advances in Neural Information Processing Systems}, volume~30. Curran Associates, Inc., 2017.

\bibitem{10.5555/3495724.3496298}
Jonathan Ho, Ajay Jain, and Pieter Abbeel.
\newblock Denoising diffusion probabilistic models.
\newblock In {\em Proceedings of the 34th International Conference on Neural Information Processing Systems}, NIPS '20, Red Hook, NY, USA, 2020. Curran Associates Inc.

\bibitem{ho2021classifierfree}
Jonathan Ho and Tim Salimans.
\newblock Classifier-free diffusion guidance.
\newblock In {\em NeurIPS 2021 Workshop on Deep Generative Models and Downstream Applications}, 2021.

\bibitem{lipman2023flow}
Yaron Lipman, Ricky T.~Q. Chen, Heli Ben-Hamu, Maximilian Nickel, and Matthew Le.
\newblock Flow matching for generative modeling.
\newblock In {\em The Eleventh International Conference on Learning Representations}, 2023.

\bibitem{10.1093/mnras/sty1264}
J~W Nightingale, S~Dye, and Richard~J Massey.
\newblock Autolens: automated modeling of a strong lens’s light, mass, and source.
\newblock {\em Monthly Notices of the Royal Astronomical Society}, 478(4):4738--4784, 05 2018.

\bibitem{OHare:2024nmr}
Ciaran A.~J. O'Hare.
\newblock {Cosmology of axion dark matter}.
\newblock {\em PoS}, COSMICWISPers:040, 2024.

\bibitem{Peebles2022DiT}
William Peebles and Saining Xie.
\newblock Scalable diffusion models with transformers.
\newblock {\em arXiv preprint arXiv:2212.09748}, 2022.

\bibitem{song2021scorebased}
Yang Song, Jascha Sohl-Dickstein, Diederik~P Kingma, Abhishek Kumar, Stefano Ermon, and Ben Poole.
\newblock Score-based generative modeling through stochastic differential equations.
\newblock In {\em International Conference on Learning Representations}, 2021.

\bibitem{10.5555/3298023.3298188}
Christian Szegedy, Sergey Ioffe, Vincent Vanhoucke, and Alexander~A. Alemi.
\newblock Inception-v4, inception-resnet and the impact of residual connections on learning.
\newblock In {\em Proceedings of the Thirty-First AAAI Conference on Artificial Intelligence}, AAAI'17, page 4278–4284. AAAI Press, 2017.

\bibitem{10.5555/3295222.3295349}
Ashish Vaswani, Noam Shazeer, Niki Parmar, Jakob Uszkoreit, Llion Jones, Aidan~N. Gomez, \L{}ukasz Kaiser, and Illia Polosukhin.
\newblock Attention is all you need.
\newblock In {\em Proceedings of the 31st International Conference on Neural Information Processing Systems}, NIPS'17, page 6000–6010, Red Hook, NY, USA, 2017. Curran Associates Inc.

\bibitem{1284395}
Zhou Wang, A.C. Bovik, H.R. Sheikh, and E.P. Simoncelli.
\newblock Image quality assessment: from error visibility to structural similarity.
\newblock {\em IEEE Transactions on Image Processing}, 13(4):600--612, 2004.

\bibitem{Weinberg_2015}
David~H. Weinberg, James~S. Bullock, Fabio Governato, Rachel Kuzio~de Naray, and Annika H.~G. Peter.
\newblock Cold dark matter: Controversies on small scales.
\newblock {\em Proceedings of the National Academy of Sciences}, 112(40):12249–12255, February 2015.

\bibitem{wildberger2023flow}
Jonas~Bernhard Wildberger, Maximilian Dax, Simon Buchholz, Stephen~R Green, Jakob~H. Macke, and Bernhard Sch{\"o}lkopf.
\newblock Flow matching for scalable simulation-based inference.
\newblock In {\em Thirty-seventh Conference on Neural Information Processing Systems}, 2023.

\end{thebibliography}

\appendix

\section{Training Setup}

We trained our flow-matching model using grayscale lensing images, normalized to the range $[-1, 1]$. The model operates in a latent space of dimension 512, with patch-based processing and a transformer backbone. An exponential moving average (EMA) of model weights was maintained throughout training to improve stability and sample quality. Table~\ref{tab:hyperparams} summarizes the key architectural and optimization choices.

\begin{table}[htbp]
    \centering
    \caption{Model and training hyperparameters.}
    \label{tab:hyperparams}
    \begin{tabular}{ll}
        \toprule
        \textbf{Parameter} & \textbf{Value} \\
        \midrule
        Latent dimension & 512 \\
        Patch size & 2 \\
        Model depth & 6 \\
        Attention heads & 8 \\
        Optimizer & AdamW \\
        Learning rate & $1 \times 10^{-5}$ \\
        Batch size & 128 \\
        Training epochs & 300 \\
        EMA & Enabled \\
        \bottomrule
    \end{tabular}
\end{table}

Training was performed on a single NVIDIA RTX A6000 Ada GPU using PyTorch. Each training run spanned 300 epochs with a batch size of 128. The AdamW optimizer was employed with a constant learning rate of $1 \times 10^{-5}$. All experiments were conducted in mixed precision to balance efficiency and numerical stability.

\end{document}